\documentclass{article}
\usepackage{ijcai26}

\usepackage{times}
\usepackage{soul}
\usepackage{url}
\usepackage[hidelinks]{hyperref}
\usepackage[utf8]{inputenc}
\usepackage[small]{caption}
\usepackage{graphicx}
\usepackage{amsmath}
\usepackage{amsthm}
\usepackage{booktabs}
\usepackage{algorithm}
\usepackage[switch]{lineno}

\usepackage{amssymb}
\usepackage[noend]{algpseudocode}
\usepackage[inline]{enumitem}
\usepackage{tikz}
\usepackage{multirow}
\usepackage[table]{xcolor}
\usepackage{soul}
\sethlcolor{gray!30}
\usepackage{subcaption}
\usepackage{listings}
\usepackage{microtype}
\usepackage[all]{nowidow}
\usepackage{adjustbox}
\usetikzlibrary{graphs,fit,shapes.geometric,arrows.meta,decorations.pathmorphing,backgrounds,positioning,fit,petri}
\graphicspath{{./src/}}

\newtheorem{definition}{Definition}
\newtheorem{proposition}{Proposition}

\newcommand{\defer}{\textbf{$\bigstar$}}
\newcommand{\tw}{TW$^*$}
\newcommand{\tmgtd}{\textbf{$\mathcal{M}$}}
\algnewcommand{\LineComment}[1]{\State \(\triangleright\) #1}
\newcommand{\dashline}{\raisebox{2pt}{\tikz{\draw[-,black,dashed,line width =
        1.5pt](0,0) -- (5mm,0);}}}

\newtheorem{example}{Example}
\newtheorem{theorem}{Theorem}

\title{Taming Treewidth DP with Modulators: A General Booster for Graph Heuristics}

\author{
Jialiang Li$^1$
\and
Aneta Neumann$^1$\and
Frank Neumann$^{1}$\and
Hung Nguyen$^1$\And 
Mingyu Guo$^1$\\
\affiliations
$^1$Adelaide University\\
\emails
\{j.li, aneta.neumann, frank.neumann, hung.nguyen, mingyu.guo\}@adelaide.edu.au
}

\begin{document}

\maketitle

\begin{abstract}
  {\em Treewidth} is a fundamental graph invariant that quantifies how tree-like a given graph is. It is extensively used with dynamic programming to design {\em fixed-parameter tractable} algorithms for many NP-hard graph combinatorial optimization problems. However, despite broad theoretical applicability, treewidth dynamic programming (TDP) does not scale in practice beyond graphs with very small treewidth. Rather than applying TDP as a standalone technique, in this paper, we demonstrate that TDP can serve as a broadly applicable {\em enhancer} for a wide range of graph combinatorial optimization algorithms. Our framework leverages the concept of {\em treewidth modulators}, which refer to vertex sets whose removal significantly reduces the treewidth. We further propose an empirically efficient procedure for generating such treewidth modulators.
  To enhance an algorithm $\mathcal{A}$, we use $\mathcal{A}$ to heuristically make decisions on the modulators vertices, after which the remaining decisions outside the treewidth modulators become scalable for TDP.

  To demonstrate the general applicability of our proposed framework. We experimented with three classic graph combinatorial optimization models: {\em Maximum Independent Set}, {\em Minimum Vertex Cover}, and {\em Max Cut}. We apply TDP to enhance algorithms across diverse paradigms, including {\em evolutionary search}, {\em greedy heuristics}, and {\em graph-neural-network-based heuristics}. For all combinations of optimization models and base algorithms, TDP significantly improves performance over the original methods. In many settings, TDP-enhanced greedy heuristics are {\em competitive with}, and {\em sometimes clearly outperform}, state-of-the-art commercial solvers.
\end{abstract}

\section{Introduction}

 {\it Treewidth (TW)} is a prominent graph invariant that measures the similarity
between a given graph and an exact tree, whose value can be calculated via
a {\it tree decomposition} process.
Many NP-hard graph combinatorial optimization problems become easy for special graph instances with small treewidths.
According to the well-known {\it Courcelle's Theorem}~\cite{Courcelle1990TheMS}, there is
a general way to construct {\it treewidth dynamic programs (TDP)}~\cite{downey2013fundamentals} that can optimally solve a wide spectrum of hard graph problems in {\it linear time} for graphs with bounded treewidths. These problems include but are not limited to {\it independent set}, {\it vertex cover}, {\it 3-coloring}, {\it TSP} and etc~\cite{Bodlaender1988DynamicPO}. These problems are referred to as {\it fixed-parameter tractable} parameterized by treewidth. Their time complexities are of the form $O(f(TW)\cdot n)$, where $f(TW)$ is a constant factor generally exponential in $TW$ when it comes to NP-hard problems.\footnote{For $O(f(TW)\cdot n)$, if $f$ is polynomial in $TW$, then the whole expression becomes polynomial in $n$ as $TW<n$.}
That is,
despite its general applicability and the impressive ``linear'' time complexity, TDP is practically {\bf not} scalable, unless we restrict to special graph instances with tiny treewidths.

We take the {\em Maximum Independent Set (MIS)} as an example.
There exists a treewidth dynamic program with complexity $O(2^{TW}n)$. Going by real-world graphs~\cite{Maniu2019AnES}, for a power network with treewidth $10$, the above TDP is certainly scalable. For a road network with treewidth $50$, the above TDP becomes not scalable due to the large constant factor $2^{50}$.
Again, it is {\em expected} that TDP is not scalable, as MIS is, after all, an NP-hard problem.

In summary, the advantage of TDP is its broad applicability and its disadvantage is its limited practically scalability.
Instead of applying TDP as a standalone technique, we propose to apply TDP to enhance existing graph optimization algorithms.
We employ
the concept of {\it treewidth modulator}. It is a vertex
deletion problem~\cite{Lewis1980TheNP}, with the goal of deleting
as few vertices as possible to bring down the treewidth to
below a desired threshold (i.e., to a threshold that makes
TDP scalable).
Our usage of treewidth modulator certainly does not involve actually deleting any vertices, as otherwise that would be modifying the original optimization model and/or the solution space.
We use the treewidth modulator as a {\it backdoor}~\cite{Williams2003BackdoorsTT}, which is a concept borrowed from SAT (Boolean satisfiability). For SAT, it is known that {\it hard instances can become much easier by fixating assignments to a small subset of variables}.
The assignments for reducing computational difficulty are called the backdoor.
We apply TDP to enhance an existing graph optimization algorithm $\mathcal{A}$ as follows. (Our approach is broadly applicable to many optimization models, which is why our presentation intentionally avoids referencing a specific model.
The readers are welcome to adopt  {\it Maximum Independent Set} as an example, if it aids in understanding.)
We use $\mathcal{A}$
to make decisions within the treewidth modulator.
We refer to these decisions as the {\it advice string}, which serves to significantly reduce the TDP search space (i.e., it is the ``backdoor'' to TDP).
We propose a modified version of TDP that takes as input an advice string, and focuses on generating the decisions outside of the treewidth modulator, which is scalable (as it is effectively equivalent to working on the remaining graph after vertex deletion).
Essentially, our modified version of TDP produces the {\em optimal solution {\bf conditional on} the advice string}.
The above interpretation also shows that for any existing algorithm $\mathcal{A}$, the TDP enhanced version of $\mathcal{A}$ never performs worse than $\mathcal{A}$ itself in terms of solution quality.

\subsection{Summary of contributions}
\begin{itemize}
  \item We propose a general framework on using TDP to enhance existing graph combinatorial optimization algorithms. We experimented on three different graph
        optimization models, including
          {\it Maximum Independent Set}, {\it Minimum Vertex Cover}, and {\it Max Cut}.
        We apply TDP to enhance algorithms following
        a variety of paradigms, including
          {\it (1+1)EA evolutionary algorithm}, {\it greedy}, and {\it graph neural networks based heuristics}.
        For all combinations of optimization models and existing algorithms, TDP significantly enhances the existing algorithms.

  \item Our framework builds on the efficient generation of treewidth modulators. Optimal treewidth modulator generation is known to be notoriously difficult. We propose a scalable variant called {\it Optimal Treewidth Modulator on a Given Tree Decomposition}. We prove that our variant is still NP-hard, but it is empirically cheap, costing only 1 second on average in our experiments.

  \item For easy graph datasets, TDP enhanced (1+1)EA almost always reaches the optimal solution, and it is significantly superior to standalone (1+1)EA in terms of wall-clock time and number of iterations before hitting optimality.

  \item  In many scenarios with hard graph datasets, TDP enhanced greedy heuristics are {\it competitive with}, and {\it sometimes clearly outperform}, the solutions by {\sc Gurobi}. Our {\it pure Python} implementation manages to reach {\it better solutions} while using {\it less time} than {\sc Gurobi}.

  \item  TDP enhanced graph neural networks based heuristics
        outperform standalone graph neural networks based heuristics.
        Furthermore, we observe the following common phenomenon in training. The network quickly stagnates
        on decisions within the treewidth modulator. Effectively, we have already entered into a local optimum. Afterwards,
        there is a prolonged period of minor improvements on decisions outside the treewidth modulator.
        That is, judging by the cost function, it is decreasing and therefore {\it appears to be learning}. However, the decisions outside the treewidth modulator are not worth the training efforts as they can be optimally and efficiently derived via TDP.
        In other words, our framework can function as an early detection test to prevent such wasteful training.
\end{itemize}

\paragraph{Related Work.} Treewidth modulator is related to the {\it forbidden minor} problem, which has been extensively studied in the parameterized complexity community~\cite{Cygan2011OnTH,Baste2020HittingMO,Jansen2021VertexDP}. To the best of our knowledge, efficient algorithms for acquiring optimal treewidth modulators appear highly improbable by many conjectures~\cite{Baste2020HittingMO}. Even the fixed-parameter tractable results are based on the precondition of {\it finite graph collections}, but yet, exhibit a {\it superexponential} growth rate. The progress on treewidth modulators serves more for theoretical interests. In this paper, we do not pursue the exact calculation of treewidth modulators. Instead, we adopt a more practical approach by proposing an empirically scalable heuristic. Despite the resemblance between treewidth modulator and {\it strong backdoor}, the difference has already been clarified in \cite{Fomin2015SolvingDV}. In terms of algorithm design, other research focused on parameterized approximation scheme~\cite{Lampis2013ParameterizedAS}, which shares the same interest on accelerating treewidth dynamic programming, but via algebraic operations and non-trivial hand-crafted rules.

\section{Preliminary}

In this paper, we follow the standard notations and terminologies in graph theory and parameterized complexity. Contents marked by (\underline{\underline{\bf \defer}}) are deferred to the appendix.

\paragraph{Notations.} Given a graph $G=(V,E)$, we note $|V|=n$, $|E|=m$. We define the induced subgraph by a subset $S\subseteq V$ as $G[S]$. The neighbors of a vertex or a set of vertices are denoted by $N[\cdot]$ (inclusion-wise) and $N(\cdot)$ (exclusion-wise). We define the {\it boundary} of $S$ as $\partial_G(S)=\{v\in S: \exists u\in V\setminus S, (u,v)\in E\}$.
We focus on vertex selection problems, which
involve searching for a solution within the binary combinatorial space $2^n$. We represent an arbitrary {\it assignment} within the space as $\boldsymbol{s}\in 2^n$.
Unless otherwise specified, we assume {\it maximization} objective, such as maximizing the number of vertices selected subject to feasibility.

We start by introducing a few relevant concepts from parameterized complexity:

\begin{definition}[\bf Fixed-parameter Tractable\label{def:fpt}]
  For a problem $L\subseteq \Sigma^*\times \mathbb{N}$, if there exists an algorithm $\mathcal{A}$ that can determine whether an instance $(x,k)\in \Sigma^*\times \mathbb{N}$ in time $f(k)\cdot |x|^{O(1)}$, where $f$ is a computable function, then the problem is fixed-parameter tractable (FPT) parameterized by $k$.
\end{definition}

In our context, we focusing on using the treewidth (TW) as the parameter, which is formally defined below.
We say a graph combinatorial optimization problem is FPT parameterized by TW if we have an algorithm with complexity $O(f(TW)\cdot n^{O(1)})$. That is, if the graph instances have bounded treewidths, then the algorithm is polynomial-time.

\begin{definition}[\bf Tree Decomposition~\cite{Cygan2020ParameterizedA}\label{def:tw}]
  A tree decomposition is a mapping from G to a pair $\mathcal{T}=(T, \{X_t\}_{t\in V(T)})$, where $T$ is a tree and every tree node $t\in V(T)$ corresponds to a subset of vertices from the original graph, denoted as $X_t\subseteq V(G)$. $X_t$ is often referred to as a {\bf bag}. For valid tree decomposition, we must have:
  \begin{itemize}
    \item The union of all bags contains all the vertices from the original graph: $\bigcup_{t\in V(T)} X_t = V(G)$;
    \item For any edge $(u,v)$ in the original graph, there must exist at least one bag that contains both $u$ and $v$: $\forall (u,v) \in E(G)$, $\exists t\in T$ with $\{u,v\} \subseteq X_t$;
    \item For any vertex $u$ from the original graph, all bags containing $u$ must form a subtree of $T$: $\forall u\in V(G), T_u=\{t\in V(T) | u\in X_t\}$ is a connected component of $T$.
  \end{itemize}
  The {\bf width} of a tree decomposition $\mathcal{T}=(T, \{X_t\}_{t\in V(T)})$ is defined by
  $$\max_{t\in V(T)} |X_t| - 1$$
  The optimal {\bf treewidth} of a graph $G$ is denoted by \tw, which is the minimum width of all valid tree decompositions of $G$. We denote the tree decomposition producing the minimum treewidth as $\mathcal{T}^*$.
\end{definition}

It should be noted that the computation of the minimum treewidth is proven NP-hard~\cite{Robertson1984GraphMI}, hard to approximate~\cite{Wu2014InapproximabilityOT}, and hard even for special graph classes~\cite{Bodlaender2023TreewidthIN}. For practical purposes, treewidth dynamic programming does not require the minimum treewidth. It can be based on sub-optimal treewidth
values obtained via polynomial-time tree decomposition heuristics. In our presentation, treewidth refers
to a heuristically generated value.

The following theorem proves that bounded treewidth implies
linear-time algorithms for many problems.

\begin{theorem}[\bf Courcelle's Theorem~\cite{Courcelle1990TheMS}\label{def:coucelle}]
  Given a monadic second order logic (MSO) formula $\varphi$, and a graph instance $G$. Then, there exists a linear-time algorithm (in time $f(|\varphi|, k)$) to decide if $G$ satisfies $\varphi$ if TW(G) is bounded by $k$.
\end{theorem}

Problems such as {\it 3-coloring}, {\it Hamiltonian cycle}, {\it independent set}, and {\it vertex cover} are all expressible by MSO. According to the above theorem, they
all are linear-time solvable for graph instances with bounded treewidths. There is also a systematic way to construct treewidth dynamic programs for these problems.
Theoretically, this sounds nice, but practically, direct application of TDP is often not scalable due to the inevitable exponential constant factor in the treewidth.
We apply TDP as a broadly applicable tool to enhance existing graph optimization algorithms. Our framework employs a similar concept to treewidth modulator.

\begin{definition}[\bf Treewidth Modulator~\cite{Cygan2011OnTH}]\label{def:tm}
  Let $\eta\ge 0$ be an integer and $G$ be a graph. A set $\mathcal{M}_{\eta}\subset V(G)$ is called an $\eta$-treewidth modulator in $G$ if $TW^*(G\setminus \mathcal{M}_{\eta})\le \eta$. We use $\mathcal{M}^*_{\eta}(G)$ to denote the optimal treewidth modulator to a target width value $\eta$ with the minimum size (i.e., we want to minimize the number of vertices deleted $|\mathcal{M}_{\eta}|$).
\end{definition}

We use $\mathcal{M}^*$ to denote the task of generating the optimal treewidth modulator $\mathcal{M}^*_{\eta}(G)$, which
is NP-hard due to its hereditary property~\cite{Lewis1980TheNP}. Its hardness
can also be easily illustrated via the following example.

\begin{example}\label{exp:tm}
  When $\eta=1$, $\mathcal{M}^*$ is equivalent to {\it feedback vertex set}.
  When $\eta=0$, $\mathcal{M}^*$ is equivalent to {\it vertex cover}.
\end{example}

Despite a long series of existing theoretical works on $\mathcal{M}^*$, such as \cite{Baste2020HittingMO}, there do not exist scalable exact algorithms, approximation algorithms, or
even practically useful bounds.
In this paper, we propose a variant of $\mathcal{M}^*$ called
  {\em Treewidth Modulator on a Given Tree Decomposition}.\footnote{
  Our proxy problem is on deleting vertices. A variant on deleting edges from a given tree decomposition has been previously studied in the field of bioinformatics~\cite{Marchand2021TreeDR}.} Our variant was designed for {\em practical scalability} -- it is still NP-hard, but its exact solution takes only 1 second on average in our experiments.

\begin{definition}\label{def:TMGTD}
  \textbf{(Treewidth Modulator on a Given Tree Decomposition)}
  Given a graph $G$ and a tree decomposition of $G$ denoted as $\mathcal{T}=(T, \{X_t\}_{t\in V(T)})$ and a parameter $k$. We simply delete vertices from the existing bags (the $X_t$) and do not regenerate the tree structure $T$. That is, after deleting $Y\subseteq V(G)$, the resulting tree decomposition is $\mathcal{T}'=(T, \{X_t\setminus Y\}_{t\in V(T)})$ and the resulting treewidth is $\max_{t\in V(T)} |X_t\setminus Y| - 1$.
  We want to decide whether there exists a vertex set $Y\subseteq V(G), |Y|\le k$, whose deletion will reduce the width of $\mathcal{T}'$ to at most $\eta$.
\end{definition}

We use $\tmgtd$ to denote the above variant.\footnote{We abuse $\mathcal{M}$ for both treewidth modulator and the problem of computing treewidth modulator.} The idea behind it is simple. Instead of regenerating tree decomposition after vertex deletion, we reuse the same tree decomposition by simply removing vertices from its existing bags.
We first show that the deleting vertices from a tree decomposition's existing bags indeed results in a valid tree decomposition.

\begin{proposition}[\bf \defer]
  Given a graph $G$ and a tree decomposition of $G$ denoted as $\mathcal{T}=(T, \{X_t\}_{t\in V(T)})$, where $T$ is a tree and every tree node $t\in V(T)$ corresponds to a subset of vertices from the original graph, denoted as $X_t\subseteq V(G)$.

  For any set of vertices $Y\subseteq V(G)$, after deleting $Y$ from $\mathcal{T}$ without regenerating the tree decomposition, the result $\mathcal{T}'=(T, \{X_t\setminus Y\}_{t\in V(T)})$ is still a valid tree decomposition.
\end{proposition}

Next, we claim that even though {\tmgtd} avoids regenerating tree decompositions, it remains NP-complete.

\begin{theorem}[\bf \defer]
  Treewidth modulator on a given tree decomposition is {\sc NP-complete}.
\end{theorem}

In our experiments, we use the polynomial-time {\sc min-degree} heuristic
to generate the initial tree decomposition and apply mixed integer programming to remove the minimum set of vertices
from the initial tree decomposition's existing bags ($\defer$). The above is a fast heuristic for $\mathcal{M}$.

\section{Technical Description of Framework}

We start by describing our {\it core routine}, which
applies TDP to enhance an existing algorithm specified
as a subroutine. We describe the {\it (1+1)EA evolutionary algorithm} subroutine and the {\it $\epsilon$-greedy} subroutine in this section. We defer the {\it graph neural networks (GNN) based heuristics} to the experiments.
In our presentation, we use $P$ to denote the optimization problem with solution space $2^n$. Without loss of generality, we assume that $P$ has a {\it maximization} objective.

\paragraph{Core Routine.}
There are no known efficient algorithms for
generating the optimal treewidth modulator. Our more scalable variant of treewidth modulator on a given tree decomposition ($\mathcal{M}$) offers an empirically scalable way to generate a treewidth modulator that suffices for our TDP application.
While treewidth modulators are defined as vertex sets whose deletion would reduce the treewidth of the remaining graph, when we apply a treewidth modulator, we do not {\it actually delete} any vertices, but we rely on the existing algorithm $\mathcal{A}$ that we aim to enhance (i.e., the subroutine we plug in) to assign values to $v\in\mathcal{M}$ from a search space of $2^{|\mathcal{M}|}$. The assignments are referred to as $\mathcal{A}$'s {\it advice}, denoted by $s$. Via TDP, we can efficiently derive the optimal overall solution {\it conditional} on advice $s$. To illustrate this idea, imagine that we are processing a specific bag $\mathcal{B}$ from the tree decomposition. Typically, TDP involves enumerating all feasible states/assignments of $\mathcal{B}$. For example, suppose $|\mathcal{B}|=4$. Our initial understanding of the optimal assignments can be described as the string \texttt{<?,?,?,?>}, representing that the optimal assignments are completely unknown. TDP needs to enumerate $2^4$ states. Suppose an {\it advice} is provided, we can settle some of the $?$s. Assume $\mathcal{A}$ advised that the first coordinate must be $1$ and the third coordinate must be $0$, i.e., our understanding of the optimal assignments becomes \texttt{<1,?,0,?>}. Consequently, we only have $2^2$ states to enumerate. Thus, given an advice string describing the decisions within the treewidth modulator, TDP is effectively focusing on the decisions outside the treewidth modulator, as if the treewidth modulator vertices have been deleted. Our core routine's pseudo code is provided as follows. We follow the standard description of treewidth dynamic program,\footnote{Please refer to Chapter 7.1-7.3 of \cite{Cygan2020ParameterizedA} for a detailed description of treewidth dynamic programming.} only slightly modified to accommodate advice.

\begin{algorithm}[!ht]
  \caption{Treewidth Dynamic Programming with {\it advice}}\label{alg:TDPA}
  \textbf{Input}: $G=(V,E)$, $\mathcal{T}_w=(T, \{X_t\}_{t\in V(T)})$: an arbitrary tree decomposition with width $w$, $\mathcal{M}_{\eta}$: a modulator to $\mathcal{T}_w$ with target width $\eta$, $\boldsymbol{s}\subseteq\mathcal{M}$: an {\it advice string} (subset) given by external subroutine $\mathcal{A}$, $P$: a maximization problem\\
  \textbf{Output}:  {\sc OPT} to P
  \begin{algorithmic}[1]
    \State Let $C$, $D$ be two tables.
    \For{$t\in V(\mathcal{T})$ in a {\it bottom-up} manner}
    \State $\mathcal{X}_p \leftarrow \{k | k\in V(T) \text{ and } k\in \text{Parent}(t)\}$
    \State $\mathcal{X}_c \leftarrow \{j| j\in V(T) \text{ and } j\in \text{Children}(t)\}$
    \Statex\textit{Apply advice here}
    \State $\mathcal{F}_t \leftarrow \{ $x$ \mid x\subseteq X_t \setminus \boldsymbol{s} \text{ and } \text{P($x$) is satisfied}\}$
    \For{$x \in \mathcal{F}_t$}
    \Statex\it Complete state with {\it advice}
    \State $x\leftarrow x\cup \boldsymbol{s}$
    \Statex\it Merge states from children
    \State $C(t, x\cap X_{j})=|x| + \sum_{j\in \mathcal{X}_c} (D(j, t, x\cap X_{j}) - |x\cap X_{j}|)$
    \Statex\it Upload states to parent
    \State $D(t, k, x\cap X_k) = \max_{k\in \mathcal{X}_p} C(t, x\cap X_k)$
    \EndFor
    \EndFor
    \State $OPT \leftarrow \max_{x\in \mathcal{F}_{root}} C(root, x)$
    \State \textbf{return} {\sc OPT}
  \end{algorithmic}
\end{algorithm}

\paragraph{(1+1)EA Evolutionary Algorithm Subroutine.}
We use (1+1)EA to search over advice assignments on the modulator $\mathcal{M}$. The algorithm maintain an incumbent $x\in\left\{0,1\right\}^{\left|\mathcal{M}\right|}$. Each iteration generates an offspring $y$ by mutating $x$: for every coordinate $i$, the bit $x_i$ is flipped independently with probability $1/\left|\mathcal{M}\right|$. The selection rule is elitist: $y$ replaces $x$ if it achieves at least the same fitness value.

The fitness function $f$ is defined through conditional optimization. For an advice vector $x$, we fix the decisions of vertices in $\mathcal{M}$ according to $x$, and compute an optimal completion over $V\setminus\mathcal{M}$ that inherits these fixed decisions. The resulting objective value is $f(x)$. This evaluation is performed by the treewidth DP with advice (Alg.~\ref{alg:TDPA}), which makes fitness computation scalable after modulator restriction. The complete (1+1)EA procedure is given in Alg.~\ref{alg:EA-TDPA}.

\begin{algorithm}[!ht]
  \caption{(1+1)EA Subroutine}\label{alg:EA-TDPA}
  \textbf{Input}: modulator $\mathcal{M}\subseteq V$ with $\left|\mathcal{M}\right|=m$; fitness function $f:\{0,1\}^m \rightarrow \mathbb{R}$; iteration (or time) budget $B$ \\
  \textbf{Output}: the best advice $x\in\left\{0,1\right\}^m$ found
  \begin{algorithmic}[1]
    \State Sample $x \sim \mathrm{Unif}(\{0,1\}^m)$; $f_x \leftarrow f(x)$
    \For{$\ell = 1$ \textbf{to} $B$}
    \State $y \leftarrow x$
    \For{$i = 1$ \textbf{to} $m$}
    \If{$\mathrm{Bernoulli}(1/m) = 1$}
    \State $y_i \leftarrow 1 - y_i$
    \EndIf
    \EndFor
    \State $f_y \leftarrow f(y)$
    \If{$f_y \ge f_x$}
    \State $x \leftarrow y$; $f_x \leftarrow f_y$
    \EndIf
    \EndFor
    \State \textbf{return} $x$
  \end{algorithmic}
\end{algorithm}

\paragraph{$\epsilon$-greedy with Voting Subroutine.}
The second class of algorithms we investigate is {\em greedy}. For each problem $P$, we instantiate a problem-specific greedy heuristic $\mathcal{H}_p$: {\em max-degree} for {\sc MVC}, {\em vertex-flipping} for {\sc MC}, and {\em min-degree} as an effective heuristic for {\sc MIS}.

We augment $\mathcal{H}_p$ with an {\em exploration} rate $\epsilon$, interpreted as the probability of taking a non-greedy step rather than the greedy one. Algorithm~\ref{alg:H-TDPA} runs $\mathcal{H}_{p}(G,\epsilon)$ for $R$ trails to generate solutions $x^{\left(r\right)}$, then converts each trail into an advice bitvector $a^{\left(r\right)}\in\left\{0,1\right\}^m$ by restricting to the modulator , i.e., $a^{\left(r\right)}\leftarrow x^{\left(r\right)}\mid_{\mathcal{M}}$; each trail is scores by $\alpha^{\left(r\right)}\leftarrow obj_{P}(x^{\left(r\right)})$. The final advice is selected by either:
\begin{enumerate*}
  \item $\mathsf{WTA}$: choose $r^*\in\arg\max_{r\in\left[R\right]} \alpha^{\left(r\right)}$ and output $a^{\left(r\right)}$.
  \item $\mathsf{Maj}$: for each $i\in\left[m\right]$, set $a_i=1\iff \left|\left\{r\in\left[R\right]\mid a_i^{\left(r\right)}=1\right\}\right|>R/2$.
\end{enumerate*}
In this way, \textsf{WTA} trusts the single best run under $obj_{P}$, while \textsf{Maj} stabilizes the modulator decisions through aggregation over repeated $\epsilon$-greedy trails.

\begin{algorithm}[!ht]
  \caption{$\epsilon$-Greedy with Voting Subroutine}\label{alg:H-TDPA}
  \textbf{Input}: modulator \(\mathcal{M}\) with \(|\mathcal{M}|=m\); heuristic \(\mathcal{H}_P\); exploration rate \(\epsilon\); trials \(R\); rule \(\mathsf{Vote}\in\{\mathsf{WTA},\mathsf{Maj}\}\)\\
  \textbf{Output}: advice vector \(a\in\{0,1\}^{m}\)
  \begin{algorithmic}[1]
    \For{$r=1$ \textbf{to} $R$}
    \State \(x^{(r)} \leftarrow \mathcal{H}_P(G,\epsilon)\)
    \State \(a^{(r)} \leftarrow x^{(r)}|_{\mathcal{M}} \in \{0,1\}^{m}\)
    \State \(\alpha^{(r)} \leftarrow \mathrm{obj}_P(x^{(r)})\)
    \EndFor
    \If{$\mathsf{Vote}=\mathsf{WTA}$}
    \State \(r^\star \leftarrow \arg\max_{r\in[R]} \alpha^{(r)}\)
    \State \Return \(a^{(r^\star)}\)
    \Else
    \State Initialize \(a\leftarrow \mathbf{0}\in\{0,1\}^{m}\)
    \For{$i=1$ \textbf{to} $m$}
    \If{$|\{r\in[R]\mid a^{(r)}_i=1\}| > R/2$}
    \State \(a_i \leftarrow 1\)
    \EndIf
    \EndFor
    \State \Return \(a\)
    \EndIf
  \end{algorithmic}
\end{algorithm}

\paragraph{Correctness.}
TDPA conditions the dynamic program on an advice assignment over the treewidth modulator $\mathcal{M}$. The key correctness requirement is that this conditioning must neither exclude feasible completions that remain consistent with the advice nor admit infeasible solutions. The required consistency mechanism is problem dependent, and we use {\sc MC} versus {\sc MIS}/{\sc MVC} to illustrate the difference.

For {\sc MC}, the advice only fixes decisions internal to $\mathcal{M}$ and does not impose constraints on choices outside $\mathcal{M}$. Consequently, committing to any assignment on $\mathcal{M}$ cannot introduce cross-boundary conflicts, and the DP over the remainder of the graph remains an exact optimization over all feasible completions.

Fro {\sc MIS} and {\sc MVC}, in contrast, decisions on $\mathcal{M}$ necessarily constrain vertices outside $\mathcal{M}$ through adjacency. We therefore enforce boundary consistency via $\partial_{\mathcal{M}}$, the set of vertices in $\mathcal{M}$ that have neighbors outside $\mathcal{M}$. For {\sc MIS}, selecting $v\in\partial_{\mathcal{M}}$ forces all neighbors in $N(v)$ to be non-selectable. For {\sc MVC}, excluding $v\in\partial_{\mathcal{M}}$ forces all neighbors $v\in N(v)$ to be included. These implications ensure that the DP enumerates exactly the feasible completions consistent with the advice in the next stage.

Finally, for randomized algorithms like (1+1)EA, the advice itself maybe locally infeasible within $\mathcal{M}$ (e.g., selecting adjacent vertices for {\sc MIS}). Such conflicts are resolved by a simple randomized repair step that restores feasibility on $\mathcal{M}$ before invoking TDPA.

\section{Experiments}

We evaluate on {\it Maximum Independent Set} ({\sc MIS}),
  {\it Minimum Vertex Cover} ({\sc MVC}), and {\it Max Cut} ({\sc MC}). We apply TDP to enhance {\it (1+1)EA}, {\it greedy}, and {\it graph neural networks}.
For every combination of optimization models and existing algorithms, our TDP enhanced versions significantly improve over the original algorithms. In many scenarios, TDP-enhanced greedy achieve solutions that are comparable to, and sometimes clearly outperform, those obtained by {\sc Gurobi}, a state-of-the-art commercial solver.
It is worth emphasizing that while occasionally surpassing {\sc Gurobi} is an {\it impressive achievement, especially considering that our implementation is pure Python}, our primary experimental goal is to demonstrate that TDP can be used as a broadly applicable enhancement tool.

\paragraph{Dataset.}
We construct two benchmark sets, {\sc easy} and {\sc hard}, based on treewidth. {\sc Easy} consists of real-world graphs, e.g., from {\em fluid dynamics}, {\em road networking} and others that typically have relatively low treewidth. {\sc Hard} contains synthetic instances generated from the frequently cited models~\cite{Ahn2020LearningWT,Sun2023DIFUSCOGD} {\em Erd\H{o}s--R'enyi} (ER) and {\em Barab'asi--Albert} (BA) models, configured to yield moderately large treewidth so that commercial solvers such as {\sc Gurobi} struggle to obtain high-quality solutions within a limited time budget; we generate {\em 100} instances per model. We exclude near-clique graphs (clique number close to $|V|$), since they usually have treewidth linear to $|V|$ and are thus uninformative for treewidth dynamic programming. Full instance statistics and sources are deferred ({\bf\defer}) to the appendix.

\paragraph{Baselines and Configurations.}
We evaluate baseline algorithms with the TDPA-enhanced counterparts under our framework. As a reference point, we include {\sc Gurobi}, which typically yields strong feasible solutions for combinatorial optimization even when it does not complete the optimality proof. To ensure a fair comparison, we restrict {\sc Gurobi} to a single thread, consistent with all other methods in this paper. We report results at checkpoints of $\left\{30,60,300,600\right\}$ seconds under two settings: the default configuration and \texttt{MIPFocus=1}, which prioritizes finding high-quality feasible solutions.

For heuristic baselines, we integrate evolutionary algorithm, greedy heuristics and GNN models into our framework, demonstrating broad composability across various paradigms.

\emph{Greedy heuristics.} We tune $\epsilon$ in the range of $\left\{0,0.01,0.05\right\}$ as exploration ratio on both MIS and MVC. For MC, we use vertex-flipping, here $\epsilon$ controls the refinement budget (number of iterations), and we set it in $\left\{0, 1000, 5000\right\}$. Details (\defer) are deferred to appendix.

\emph{GNN baselines.} We train a simple GNN (\defer) by unsupervised minimization of the QUBO objective (\defer). Given GNN advice, TDPA computes the optimal completion consistent with that advice.

All methods have 10 runs with 10 different seeds per instance; additional details can be found in appendix.

\paragraph{Decompositions and Modulators.}
Prior results~\cite{Bannach2017JdrasilAM,Maniu2019AnES} indicate that {\em min-degree} (\defer) heuristic often produces {\em near-optimal} treewidth in practice and remains competitive with more sophisticated treewidth algorithms. We therefore apply this heuristic to construct initial tree decompositions. And we defer the details of this heuristic to appendix. For the modulator $\mathcal{M}$, we formulate the computation as a mixed-integer programming (\defer), since in practice the MIP takes about {\em one second on average} to obtain $\mathcal{M}$. Although the optimality gap between $\mathcal{M}$ and $\mathcal{M}^*$ can in principle be large, tightening $|\mathcal{M}|$ is not our focus; any reduction in $|\mathcal{M}|$ {\em will only} strengthen our results.

\begin{figure}[!ht]
  \centering
  \includegraphics[width=1.0\linewidth]{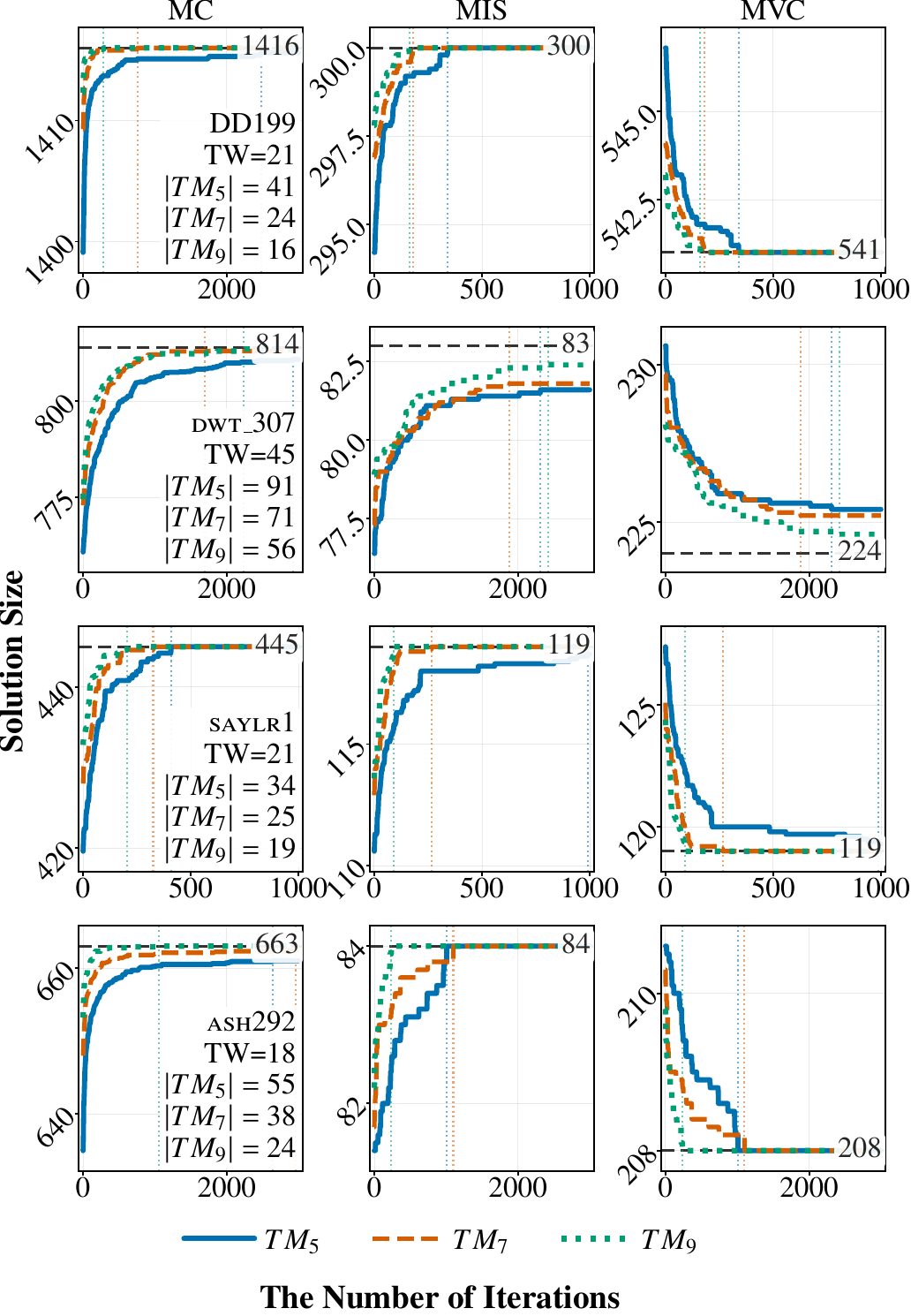}
  \caption{
    Colored vertical lines mark the {\it average first hitting time --- the first moment the best solution appears for each curve}. The \protect\dashline indicates the {\sc Gurobi} solutions. For presentation, once the curves converge to a single value, we omit the rest steps because they are not informative. The instance name and TW are shown above TM in each legend group. More results (\defer) are deferred to the appendix.
  }\label{fig:ea-tdpa}
\end{figure}

\subsection{Results \& Analysis}

\paragraph{On Randomized Algorithm.}
Across all instances and objectives (MC/MIS maximization and MVC minimization), Figure~\ref{fig:ea-tdpa} highlights two consistent effects: incorporating $\mathcal{M}$ reduces stagnation, and smaller modulators typically yield stronger gains. We denote $TM_\tau$ as the modulator computed by $\mathcal{M}$ at target treewidth $\tau$ (larger $\tau\Rightarrow$ smaller $|TM_\tau|$). We vary the target treewidth $\tau\in\{5,7,9\}$; as the legends indicate, a larger $\tau$ produces a smaller modulator, e.g., $|TM_\textrm{5}|>|TM_\textrm{7}|>|TM_\textrm{9}|$. Under a much smaller budget (3,000 vs.\ 1 million iterations), the modulator-guided runs reach high-quality solutions much earlier: the annotated first-hitting times occur sooner and the trajectories improve more steadily, with $TM_\textrm{9}$ most often leading throughout the run. Additionally, the smaller-$|\mathcal{M}|$ curves more frequently approach, and in several cases match the {\sc Gurobi} reference line, while larger-$|\mathcal{M}|$ runs exhibit longer plateaus, whereas leveraging $|\mathcal{M}|$ substantially increases the chance of attaining the global optimum. The table echoes this effect quantitatively: for many instances (e.g., DD199 and power-685-bus across MC/MIS/MVC), EA+TDPA attains the {\sc Gurobi}-optimal objective values whereas $(1+1)$EA remains far below/above them, consistent with the baseline being trapped on suboptimal plateaus. While the smaller modulator is preferable in best objective values, the first-hitting times in seconds can exhibit a trade-off (e.g., dwt\_307), and there are instances where the baseline is competitive under our settings (e.g., can\_161), suggesting room for adaptive choices of $\tau$.

\begin{table}[!ht]
  \centering
  \setlength\tabcolsep{0.1ex}
  \renewcommand{\arraystretch}{1.2}
  {\huge\scshape
    \begin{adjustbox}{width=\columnwidth}
      \begin{tabular}{lccccc}
        \toprule
        { {\sc Instances}}
         & {{\sc (1+1)EA}}
         & {{\sc Ea+Tdpa}{\bf (5)}}
         & {{\sc Ea+Tdpa}{\bf (7)}}
         & {{\sc Ea+Tdpa}{\bf (9)}}
         & {{\sc Gurobi}}           \\
        \midrule

        \multirow{3}{*}{DD199}
         & 1393.6 (553.52)
         & 1415.6 (230.77)
         & 1416.0 (80.89)
         & {\bf 1416.0 (68.32)}
         & {\bf 1416}               \\

         & 273.0 (604.30)
         & 300.0 (23.66)
         & {\bf 300.0 (16.55)}
         & 300.0 (16.70)
         & {\bf 300}                \\

         & 568.0 (622.84)
         & 541.0 (23.36)
         & {\bf 541.0 (16.03)}
         & 541.0 (16.05)
         & {\bf 541}                \\
        \midrule

        \multirow{3}{*}{dwt\_307}
         & 802.3 (82.94)
         & 810.8 (152.70)
         & 813.0 (291.82)
         & {\bf 813.5} (763.10)
         & {\bf 814}                \\

         & 77.4 (255.50)
         & 81.6 (24.59)
         & 81.8 (52.01)
         & {\bf 82.4} (100.02)
         & {\bf 83}                 \\

         & 229.6 (258.66)
         & 225.4 (24.69)
         & 225.2 (43.19)
         & {\bf 224.6} (71.79)
         & {\bf 224}                \\
        \midrule

        \multirow{3}{*}{\parbox{1.5cm}{ia-enron-only}}
         & 422.9 (23.03)
         & 425.5 (36.28)
         & 426.2 (31.29)
         & {\bf 426.5} (170.50)
         & {\bf 427}                \\

         & 56.9 (65.66)
         & 57.0 (2.00)
         & {\bf 57.0 (0.39)}
         & 57.0 (0.88)
         & {\bf 57}                 \\

         & 86.1 (66.11)
         & 86.0 (2.11)
         & {\bf 86.0 (0.34)}
         & 86.0 (0.66)
         & {\bf 86}                 \\
        \midrule

        \multirow{3}{*}{\parbox{1.5cm}{power-685-bus}}
         & 1039.9 (280.53)
         & 1069.6 (98.78)
         & 1069.8 (96.26)
         & {\bf 1070.0} (173.46)
         & {\bf 1070}               \\

         & 297.1 (436.09)
         & 313.0 (25.24)
         & 313.0 (7.51)
         & {\bf 313.0 (4.38)}
         & {\bf 313}                \\

         & 387.9 (444.00)
         & 372.0 (25.86)
         & 372.0 (7.31)
         & {\bf 372.0 (4.09)}
         & {\bf 372}                \\
        \midrule

        \multirow{3}{*}{saylr1}
         & 440.5 (5.91)
         & {\bf 445.0 (13.16)}
         & 445.0 (16.60)
         & 445.0 (33.21)
         & {\bf 445}                \\

         & 110.9 (139.41)
         & 119.0 (11.19)
         & {\bf 119.0 (8.21)}
         & 119.0 (11.23)
         & {\bf 119}                \\

         & 127.1 (142.74)
         & 119.0 (11.30)
         & {\bf 119.0 (7.81)}
         & 119.0 (10.45)
         & {\bf 119}                \\
        \midrule

        \multirow{3}{*}{ash292}
         & 655.6 (162.55)
         & 661.0 (104.27)
         & 662.5 (304.59)
         & {\bf 663.0} (245.65)
         & {\bf 663}                \\

         & 81.2 (230.45)
         & 84.0 (16.69)
         & 84.0 (20.92)
         & {\bf 84.0 (11.34)}
         & {\bf 84}                 \\

         & 210.8 (234.57)
         & 208.0 (16.33)
         & 208.0 (17.59)
         & {\bf 208.0 (8.22)}
         & {\bf 208}                \\
        \midrule

        \multirow{3}{*}{can\_161}
         & {\bf 450.4 (2.33)}
         & 448.0 (29.48)
         & 448.8 (39.56)
         & 448.8 (109.05)
         & {\bf 456}                \\

         & {\bf 40.2} (60.69)
         & 40.0 (2.68)
         & 40.0 {\bf (1.88)}
         & 40.1 (7.12)
         & {\bf 41}                 \\

         & {\bf 120.8} (62.19)
         & 121.0 (2.88)
         & 121.0 {\bf (1.79)}
         & 120.9 (6.01)
         & {\bf 120}                \\
        \midrule

        \multirow{3}{*}{rdb200l}
         & 441.4 (109.44)
         & {\bf 460.0 (11.66)}
         & 460.0 (22.35)
         & 460.0 (77.16)
         & {\bf 460}                \\

         & 94.2 (108.77)
         & {\bf 100.0 (7.65)}
         & 100.0 (12.01)
         & 100.0 (28.57)
         & {\bf 100}                \\

         & 105.8 (108.43)
         & {\bf 100.0 (7.73)}
         & 100.0 (11.49)
         & 100.0 (26.34)
         & {\bf 100}                \\
        \midrule
      \end{tabular}
    \end{adjustbox}
  }
  \caption{For each $\cdot(\cdot)$ pair, the first value is {\it the average best solution size}, the value in ($\cdot$) is {\it the average first hitting time of the best solution} in seconds. (5), (7) and (9) are different {\it target treewidth} values. For the three rows in each cell, from top to the bottom, are the size of MC, MIS, MVC, respectively.} \label{tab:ea-tdpa table}
\end{table}

\paragraph{On Greedy Heuristics.}
Table~\ref{tab:mis/mvc} summarizes the MIS/MVC results (Results on MC(\defer) has been deferred to the appendix), and reveals two patterns. First, $\epsilon$-greedy baselines are insensitive and problem-specific: on both ER and BA, increasing $\epsilon$ from 0 to 0.05 consistently worsens the runtime but does not reliably improve quality, and can even worsen it and yields larger gaps on both ER and BA. Second, TDPA provides a consistent improvement on solution quality and largely removes the tuning brittleness. For MIS, TDPA yields near-solver quality: on ER, $(0.01)$-TDPA attains a gap of about $-0.20\%$ relative to the 600s {\sc Gurobi} baseline, and on BA it reaches $0.51\%$ -- both dramatically better than greedy's 2-4\% gaps. For MVC, TDPA reduces the gap roughly by half compared to greedy, e.g., BA:3.78\%$\rightarrow$ 1.84\%. Compared with MIS, MVC seems more sensitive to the remaining hard structure outside the reduced-treewidth region. In terms of time, TDPA typically runs in $\sim$25-60s, which place it near the 30-60s solver budgets; in this regime it can match or exceed short-budget solver performance in several settings while remaining purely heuristic. These trends suggest that TDPA primarily improves the search landscape and robustness (replace fragile $\epsilon$ choices with stable performance). Additionally, restricting TDP to reduced-treewidth components can systematically elevate simple heuristics toward near-optimal solutions while preserving the scalability advantages that make greedy methods attractive on hard instances.

\begin{table*}[!ht]
  \centering
  \setlength\tabcolsep{3.5ex}
  \renewcommand{\arraystretch}{1.0}
  {\large\scshape
    \begin{adjustbox}{width=\textwidth}
      \begin{tabular}{rcccccc}
        \toprule
        \multirow{2}{*}{Method}                 &
        \multicolumn{3}{c}{ER}                  & \multicolumn{3}{c}{BA}                                                                                                                                                                                                                                                                                                      \\
        \cmidrule(lr){2-4}\cmidrule(lr){5-7}
                                                & Size $\uparrow/\downarrow$                                                                                                                                                           & Gap\% $\downarrow$ & Time(s) $\downarrow$ & Size $\uparrow/\downarrow$ & Gap\% $\downarrow$ & Time(s) $\downarrow$                   \\
        \midrule
        \textsc{Gurobi(30s)}                    & $715.94/1518.04$                                                                                                                                                                     & 7.85/3.93          & 30                   & $879.23/1383.30$           & 3.13/2.03          & 30                                     \\
        \textsc{Gurobi(60s)}                    & $729.36/1505.15$                                                                                                                                                                     & 6.13/3.11          & 60                   & $885.83/1376.41$           & 2.41/1.54          & 60                                     \\
        \textsc{Gurobi(300s)}                   & $760.17/1474.66$                                                                                                                                                                     & 2.16/1.10          & 300                  & $904.01/1358.79$           & 0.40/0.26          & 300                                    \\
        \textsc{*Gurobi(600s)}                  & $776.95/1458.39$                                                                                                                                                                     & 0.00/0.00          & 600                  & $907.67/1355.20$           & 0.00/0.00          & 600                                    \\
        \midrule
        \textsc{$\uparrow$Gurobi(30s)}          & $718.58/1515.68$                                                                                                                                                                     & 7.51/3.78          & 30                   & $887.17/1374.36$           & 2.26/1.39          & 30                                     \\
        \textsc{$\uparrow$Gurobi(60s)}          & $734.35/1500.19$                                                                                                                                                                     & 5.48/2.79          & 60                   & $893.09/1369.25$           & 1.61/1.03          & 60                                     \\
        \textsc{$\uparrow$Gurobi(300s)}         & $768.29/1467.18$                                                                                                                                                                     & 1.11/0.60          & 300                  & $905.50/1357.18$           & 0.24/0.15          & 300                                    \\
        \textsc{$\uparrow$Gurobi(600s)}         & $781.58/1453.98$                                                                                                                                                                     & -0.60/-0.30        & 600                  & $909.19/1353.66$           & -0.17/-0.11        & 600                                    \\
        \midrule
        \textsc{(0.0)-greedy}                   & $760.50/1515.44$                                                                                                                                                                     & 2.12/3.76          & $0.37/0.79$          & $885.90/1402.61$           & 2.40/3.38          & $0.39/0.71$                            \\
        \textsc{(0.01)-greedy}                  & $759.12/1520.55$                                                                                                                                                                     & 2.29/4.09          & $7.30/15.99$         & $883.08/1408.42$           & 2.71/3.78          & $7.73/14.31$                           \\
        \textsc{(0.05)-greedy}                  & $750.46/1541.06$                                                                                                                                                                     & 3.40/5.36          & $7.11/16.09$         & $870.04/1431.79$           & 4.15/5.35          & $7.52/14.47$                           \\
        \midrule
        \textsc{(0.0)-TDPA}                     & $767.63/1496.46$                                                                                                                                                                     & 1.20/2.54          & $24.68/25.60$        & $895.80/1380.93$           & 1.31/1.86          & $45.71/57.38$                          \\
        \cellcolor{gray!30}\textsc{(0.01)-TDPA} & $778.41$$/1494.38$         & -0.20/2.40         & $32.18/36.34$        & $903.08/1380.63$           & 0.51/1.84          & $52.43/57.64$        \\
                                                                                                                                                                                           \textsc{(0.05)-TDPA}                    & $775.68/1496.65$           & 0.16/2.50          & $42.90/47.43$        & $899.98/1381.49$           & 0.85/1.90          & $61.88/58.89$ \\
        \bottomrule
      \end{tabular}
    \end{adjustbox}
  }
  \caption{$\cdot$/$\cdot$ denotes the results for MIS/MVC. Largest improvement has been highlighted by the \hl{gray block}. $^*$ indicates the baseline used to calculate optimal gap. $\uparrow$ indicates the tuned {\sc Gurobi}. For greedy and TDPA, the values of $\epsilon$ are in ($\cdot$).} \label{tab:mis/mvc}
\end{table*}

\begin{figure}[!ht]
  \centering
  \begin{subfigure}[t]{.49\linewidth}\centering
    \hspace*{1.5em}{\bfseries MC on ER}
    \includegraphics[width=\linewidth,trim=0 20 0 0]{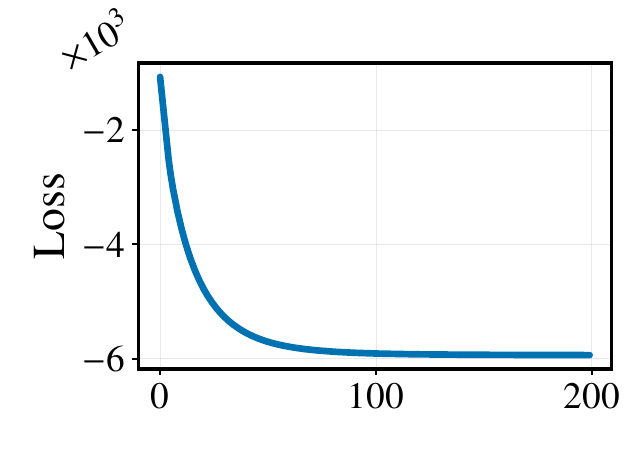}\\
    \includegraphics[width=\linewidth,trim=0 20 0 0]{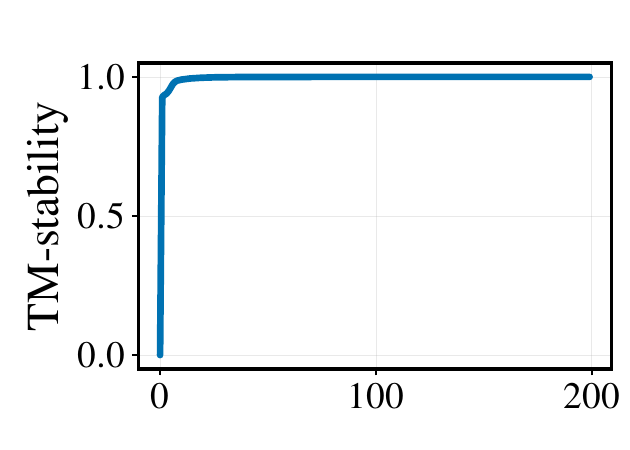}\\
    \includegraphics[width=\linewidth]{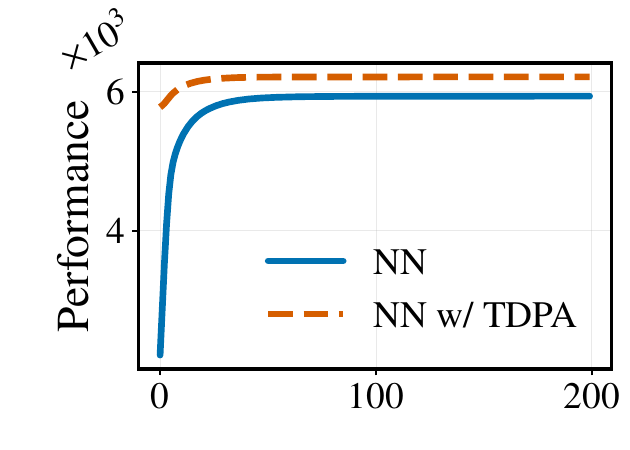}
  \end{subfigure}
  \begin{subfigure}[t]{.49\linewidth}\centering
    \hspace*{1.5em}{\bfseries MVC on ER}
    \includegraphics[width=\linewidth,trim=0 20 0 0]{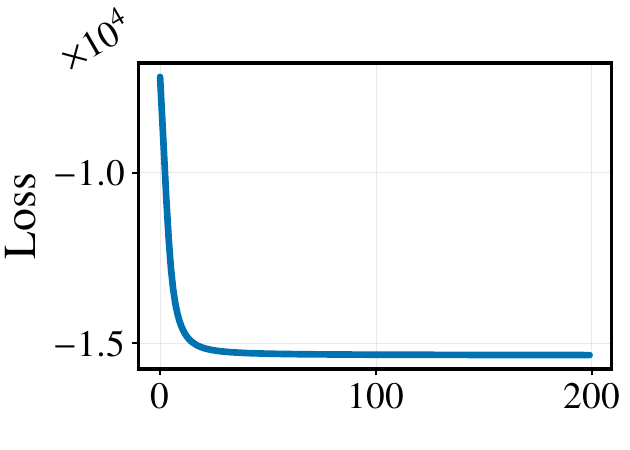}\\
    \includegraphics[width=\linewidth,trim=0 20 0 0]{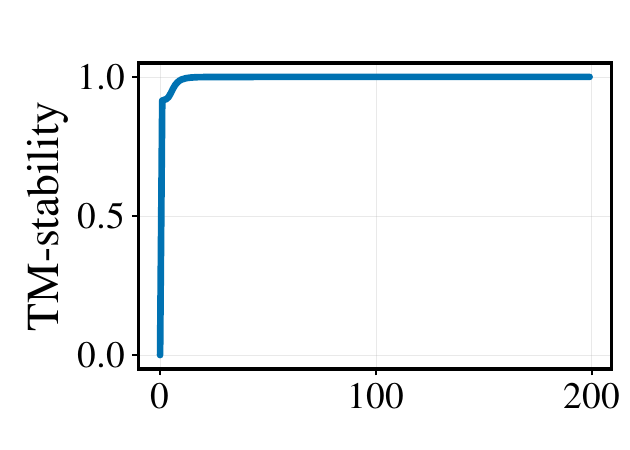}\\
    \includegraphics[width=\linewidth]{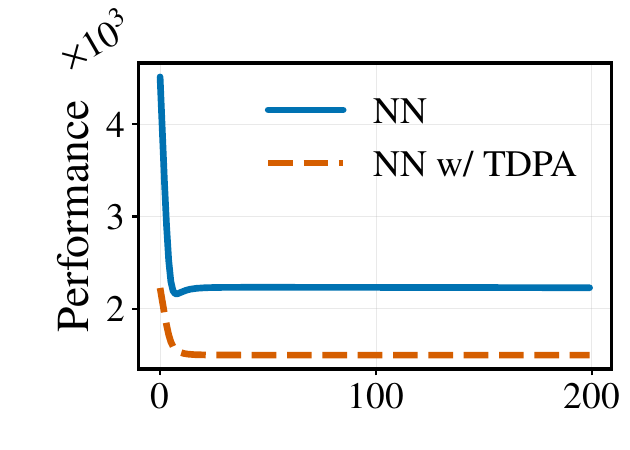}
  \end{subfigure}
  \caption{GNN training for ER. Each unit on {\it x-axis} is 50 gradient descend steps. Results for MIS and BA (\defer) are similar.}\label{fig:nn-er}
\end{figure}

\paragraph{On Graph Neural Network.}
Neural solvers for combinatorial optimization are commonly built upon GNNs with reinforcement learning~\cite{Ahn2020LearningWT} or diffusion-style sampling~\cite{Sun2023DIFUSCOGD}. In practice, they often suffer from \textit{(a) large training budgets}, \textit{(b) sensitivity to the training distribution}, and \textit{(c) limited competitiveness against strong heuristics}. We study an unsupervised alternative~\cite{Angelini2022ModernGN} that frames vertex problems as node classifications and minimize a QUBO-form objective~\cite{Glover2019QuantumBA}. Empirically, as shown in Fig.~\ref{fig:nn-er}, we find that optimization progress is not reflected by the loss: the loss decreases monotonically, yet solution quality (Performance) plateaus early, consistent with convergence to a poor local optimum; continuing training then yields diminishing returns rather than meaningful improvements.

TDPA provides a principled way to diagnose and exploit this plateau. We monitor the labels on the treewidth modulator $\mathcal{M}$ via {\em TM-stability}, defined at checkpoint $t$ as $\frac{|L_p \cap L_c|}{|\mathcal{M}|}$, where $L_t$ is the predicted label set on $\mathcal{M}$. In Fig.~\ref{fig:nn-er} (10,000 gradient-descent steps, evaluated every 50 steps), TM-stability rises to near 1.0 within the earliest checkpoints, substantially earlier than the loss and performance curves stabilize. This indicates that the model's decisions on $\mathcal{M}$ ``freeze'' quickly. This decoupling explains the stagnation: {\em subsequent updates mainly reshuffle predictions outside $\mathcal{M}$ and rarely change the globally relevant commitments captured by $\mathcal{M}$}.
From this early stabilization, TDPA uses GNN's current labels on $\mathcal{M}$ as advice to produce a high-quality feasible solution early; the orange curves track this prediction and consistently dominate the raw GNN solutions during the long plateau for MC and MIS, and yield markedly better solutions for MVC as well (where smaller the better). Consequently, TM-stability serves as an actionable {\em early-stopping signal}, and TDPA turns partial training progress into near-best solutions without paying the full training cost. Results on BA reflect the similar improvements, further confirming the effectiveness while coupling with GNNs.

\section{Conclusion}
We introduced treewidth dynamic programming with advice, a general framework for enhancing graph combinatorial optimization algorithms. TDPA integrates TDP with heterogeneous advice sources, including $(1+1)$EA, greedy heuristics, and GNN-based models. To enable this integration, we show that treewidth modulator on given tree decomposition is hard, and we propose a MIP formula that enables fast computation in practice. We then prove that the resulting advice-integrated DP remains correct. Empirically, TDPA consistently improves solution quality over the underlying methods and stays competitive with {\sc Gurobi} in first-hitting time. We further showed that TDPA provides a useful early-stage prediction, enabling earlier feedback during training and sharper improvement of downstream performance. These findings suggest that treewidth DP can be practically effective beyond small-treewidth instances when coupled with targeted advice. Future work will focus on practical methods for constructing smaller treewidth modulators to further expand the range of scalable instances.

\section*{Acknowledgements}
We would like to particularly thank \textbf{Professor Michael Fellows} and \textbf{Professor Frances Rosamond} for their valuable comments and support to this work. I am deeply grateful to \textbf{Dr. Weitong Chen} for his tremendous guidance and unwavering support. We are also profoundly thankful to \textbf{Dr. Fabien Voisin}, who consistently provided timely support with computing resources. This work was supported with resources provided by the \textbf{Phoenix HPC} service at Adelaide University. Without the support of any of these parties, this work would not have reached its current stage.

\nocite{*}
\bibliographystyle{named}
\bibliography{ijcai26_clean}

\end{document}